\documentclass[sigconf,nonacm]{acmart}

\usepackage{graphicx}
\usepackage{hyperref}
\usepackage{amsmath}
\usepackage{caption}
\usepackage{listings}
\usepackage{booktabs}
\usepackage{caption}
\usepackage{algorithm}
\usepackage{algpseudocode}

\definecolor{myred}{RGB}{180,0,0}


\begin{document}

\title{Optimizing Video Analytics
Inference Pipelines: A Case Study}

\author{Saeid Ghafouri, Yuming Ding, Katerine Diaz Chito, Jesús Martinez del Rincón, Niamh O'Connell, Hans Vandierendonck}
\email{{s.ghafouri,yding12,k.diazchito,j.martinez-del-rincon,niamh.oconnell,h.vandierendonck}@qub.ac.uk}
\affiliation{
  \institution{Queen's University Belfast}
  \city{Belfast}
  \country{United Kingdom}
}

\renewcommand{\shortauthors}{S.\ Ghafouri \emph{et al}.}

\begin{abstract}
Cost-effective and scalable video analytics are essential for precision livestock monitoring, where high-resolution footage and near-real-time monitoring needs from commercial farms generates substantial computational workloads. This paper presents a comprehensive case study on optimizing a poultry welfare monitoring system through system-level improvements across detection, tracking, clustering, and behavioral analysis modules. We introduce a set of optimizations, including multi-level parallelization, Optimizing code with substituting CPU code with GPU-accelerated code, vectorized clustering, and memory-efficient post-processing. Evaluated on real-world farm video footage, these changes deliver up to a 2$\times$ speedup across pipelines without compromising model accuracy. Our findings highlight practical strategies for building high-throughput, low-latency video inference systems that reduce infrastructure demands in agricultural and smart sensing deployments as well as other large-scale video analytics applications.
\end{abstract}



\maketitle

\section{Introduction}

Video analytics has emerged as a cornerstone technology across domains requiring automated perception and decision-making, including smart city surveillance~\cite{myagmar2023survey}, industrial automation~\cite{pavlov2024real}, autonomous vehicles~\cite{lin2022low}, and healthcare monitoring~\cite{cristina2024audio}. Recently, its application has expanded into agriculture and animal husbandry, where continuous video-based observation can provide actionable insights into welfare, health, and productivity~\cite{campbell2024computer,saraceni2024agrisort}. In particular, video analytics enables the collection of high-resolution temporal data that exceeds what is feasible through manual observation in quantity, quality and added value. 

Despite their potential, deploying large-scale video analytics systems in commercial poultry farms presents significant performance and cost challenges. A typical poultry house may contain 10,000 to 30,000 birds, and multiple houses per farm can collectively generate terabytes of high-resolution video data each week. Scaling analytics workloads involving decoding, inference, and data transfer without careful design leads to inefficient resource use and rapidly growing infrastructure costs. Optimizing video pipelines for latency is therefore critical to increase system efficiency, which in turn substantially lowers operational costs~\cite{romero2021llama,romero2022optimizing}.

This paper investigates performance bottlenecks and solutions for the
\textit{FlockFocus} pipeline, a multi-camera video analytics system developed for automated broiler chicken welfare monitoring in commercial farms~\cite{campbell2024computer}.
It analyzes high-resolution video from multiple behavioral zones including feeder, drinker, activity, and wall areas to extract metrics such as feeding frequency, locomotion, and bird density.

The system processes terabytes of video weekly across zones and houses, leading to high compute load, memory use, and data transfer.
As is common in data analytics, the system is designed using Python as the main programming language, and leveraging several highly optimized back-end libraries such as Pytorch, skimage and OpenCV. This software environment poses restrictions on the types of optimizations that can be applied.

The primary objective of our optimizations is to improve resource usage efficiency of the analytics pipelines with a view of reducing the cost of analytics. Our optimizations fall into three categories:
(i) increasing utilization of GPU and CPU computing resources through parallel execution; (ii) increasing computational efficiency of analytics by code restructuring and using efficient back-end libraries; (iii) enhancing efficiency of video input. A key lesson is that bottlenecks stem not only from algorithmic complexity, and or not centered around neural network inference. Instead, they also stem from inefficiencies in scheduling, data flow, and component interactions.

The main contributions of this work are:
\begin{itemize}
    \item A real-world case study of optimizing a multi-zone animal monitoring system, identifying architectural inefficiencies related to scheduling, I/O, and inter-stage communication in a modular pipeline.
    \item A set of system-level optimizations applied to each component module, from low-level to high-level analytics, composing the pipeline, i.e. detection, tracking, clustering, and behavior inference. Optimizations include batched and parallelized execution, GPU-accelerated post-processing, and efficient density estimation.
    \item An evaluation of the optimized pipeline on real-world farm video data, demonstrating up to a 2$\times$ reduction in processing time with minimal loss in detection and behavior classification accuracy. This corresponds to a cost reduction of over 50\% on AWS g4dn.2xlarge instances (\$0.69 to \$0.31 per run).
\end{itemize}

Although the system is designed for poultry welfare monitoring, the techniques presented here apply broadly to any video analytics pipeline involving high-resolution input, multi-stage inference, and large-scale workloads.

\begin{figure*}[t]
  \centering
  \includegraphics[width=0.95\textwidth]{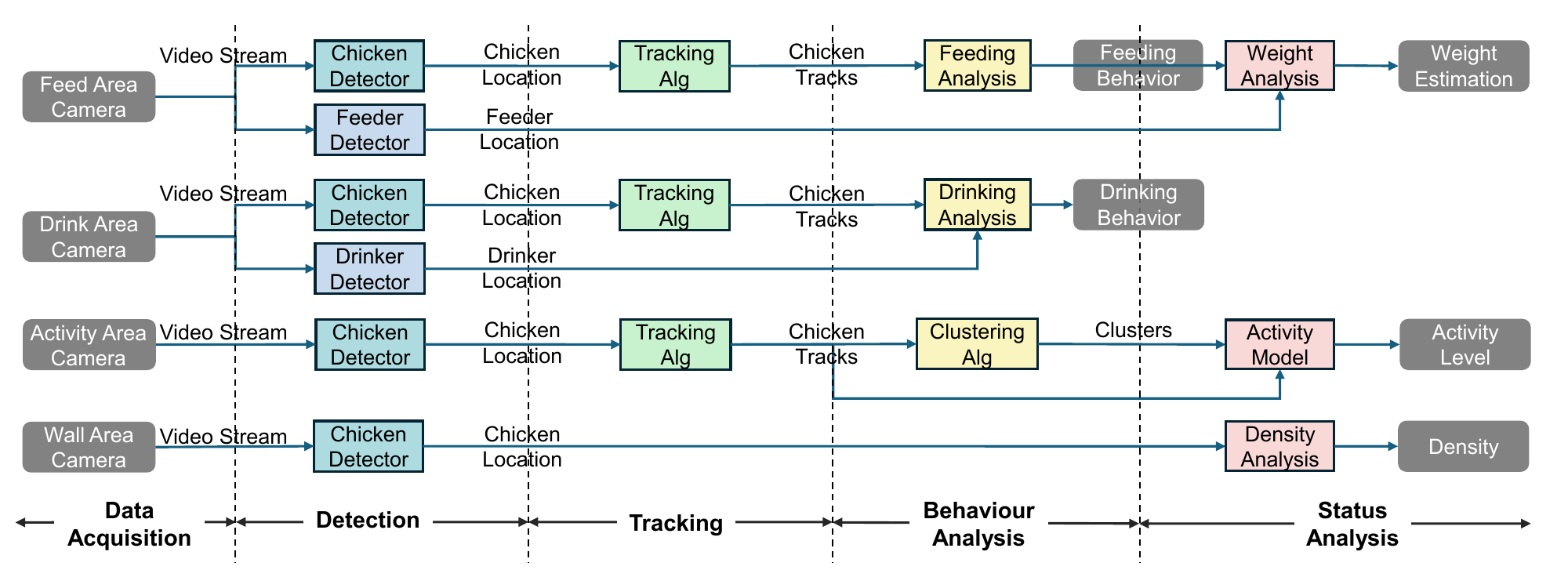}
  \vspace*{-4mm}
  \caption{FlockFocus system architecture and per-zone video pipelines. The design leverages a unified set of modules, including chicken detection, tracking, and behaviour analysis, which are adapted for each zone’s behavioural targets. Rectangles with sharp corners denote models or algorithms, grey rounded rectangles denote input data and final outputs, and the text on arrows denotes intermediate data streams.}
  \Description{Four zone-specific pipelines (feeder, drinker, activity, wall) sharing detection, tracking, and behaviour analysis modules to produce welfare metrics.}
  \label{fig:pipelines}
\end{figure*}

\section{Background: The FlockFocus System}
\label{sec:background}

In this section we first describe the building blocks of the FlockFocus system for each behavioral metric and analytical level. We will later explain the optimisations done on each of these pipelines in Section \ref{sec:optimisations}. Modern precision livestock monitoring requires robust and scalable systems capable of extracting actionable insights from high-volume, multi-camera video streams. The FlockFocus system was developed to meet these demands in commercial poultry farming, enabling automated, data-driven analysis of animal welfare at flock scale. In this section, we introduce the core architecture and deployment layout of FlockFocus, detail the zone-specific video processing pipelines that target key behavioral metrics, and describe the modular building blocks underlying each analytical stage of the system. This background provides essential context for understanding the optimization strategies in Section \ref{sec:optimisations} and empirical results presented in Section \ref{sec:evaluation}.

\subsection{System Overview and Deployment}
FlockFocus is a modular, multi-camera video analytics system designed for automated welfare monitoring in commercial poultry houses. It extracts behavioral metrics from high-resolution video recorded hourly across four key zones: feeder (feeding patterns and resource access), drinker (proximity to drinker line and engagement with the drinker), activity (locomotor activity, bird density and social clustering), and wall (bird density and social clustering). Each zone is monitored by fixed overhead cameras recording at 1080p and 30 FPS. As shown in Figure~\ref{fig:pipelines}, the system architecture supports zone-specific video-based behavioral pipelines that share a common set of analytical modules and enable scalable, automated analysis.

The pipeline is designed to record and process 5-minute video segments within each one-hour interval. 
This latency is sufficient for welfare monitoring, where timely but not instantaneous feedback is needed to detect abnormal behaviors. 
Similar notions of minute-scale delays have been adopted in video stream analytics, where near-real-time processing enables efficient event detection without the overheads of strict real-time guarantees~\cite{yadav2019vidcep}. The design supports both retrospective analysis and near real-time monitoring. The challenge faced is to reduce the financial and environmental cost of performing analytics by optimizing the processing efficiency.

\subsection{Zone-Specific Behavioral Analysis Pipelines}

Although the system shares a unified infrastructure for video capture and orchestration, each of the four zones employed a distinct processing pipeline tailored to its behavioral goals. These behavioral pipelines differ in model selection, tracking strategy, post-processing logic, and output metrics. 
An overview of these zone-specific pipelines and their core components is depicted in Figure~\ref{fig:pipelines}. The following section provides an overview of these pipelines, highlighting  differences across zones.

\noindent
\textbf{Feeder Pipeline.}  
The feeder pipeline monitors feeding behavior, by analyzing how chickens interact with feeders, and estimates individual bird weights, by observing their volume. Its broader goal is to assess feeding patterns, which can help identify welfare concerns such as competition or illness, and to monitor growth. The pipeline detects chickens and feeders in video frames, tracks chickens over time to form individual trajectories, and classifies feeding behavior based on spatial proximity to feeder lines. It then estimates bird weights using calibrated feeder dimensions and bounding box measurements~\cite{campbell2025automated}. The key outputs include feeding bouts per bird, feeding duration, and estimated weights, which together provide insight into access equity and potential health, welfare and performance issues in the flock.

\noindent
\textbf{Drinker Pipeline.}  
The drinker pipeline evaluates how chickens interact with drinker lines as a proxy for monitoring drinking behavior. It provides insight into the proportion of birds in teh drinker area that are close to the drinker line. The pipeline detects chickens and drinker lines, tracks chickens over time to obtain individual trajectories, and classifies drinking events based on spatial proximity to the nearest drinker line. Key metrics include the number of birds in the drinking area, and the percentage of these birds in close proximity to the drinker line.

\noindent
\textbf{Activity Pipeline.}  
The activity pipeline assesses movement dynamicsand social clustering of chickens in a central activity area, offering insight into behavioral diversity and welfare indicators such as inactivity or abnormal clustering behavior. The pipeline detects chickens in video frames and tracks them over time to generate individual trajectories. These tracks are used to estimate spatial clustering patterns and are analyzed by an activity module to classify each bird’s behavior as inactive, active, or highly active. The final outputs include clustering characteristics, activity labels, and merged behavior summaries that can help identify low engagement or uneven movement patterns across the flock.

\noindent
\textbf{Wall Pipeline.}  

The wall pipeline monitors flock density along the shed's wall, providing early indicators of spatial distribution issues linked to welfare concerns like heat stress. It detects chickens with a lightweight model and computes the number of birds in a region near the wall. Unlike other pipelines, it focuses on spatial density, producing per-frame counts to reflect occupancy pressure without tracking or activity classification.

\subsection{Modules of Pipelines}

FlockFocus is structured as a modular pipeline, where each analytical component such as detection, tracking, clustering, behavior analysis, and density estimation can be independently developed and optimized. The analytical level also increases, from low level features such as appearance and location, to increasingly higher features, such as identity, activity, and behaviour This section summarizes the core functionality of each module to provide context for the system-level optimizations described later.

\noindent

\textbf{Detection.} Videos are processed using a lightweight UNeXt segmentation model \cite{valanarasu2022unext} to localize birds in each frame, generating bounding boxes with confidence scores and zone-specific weights tailored to each camera setup. Importantly, model inference requires pre- and post-processing steps.
The pre-processing steps include image resizing, normalization, and conversion to tensor format. The post-processing steps vary across pipelines. When detailed objects are required, they are extracted by applying a threshold to the model's output, extracting connected components, and generating bounding boxes from the resulting masks.

The detection stage is applied in all four pipelines: feeder, drinker, activity, and wall. In the feeder and drinker pipelines, detection is paired with models for feeders and drinker lines. In the activity and feeder pipelines, detections are passed to tracking for individual trajectories. In the wall pipeline, detection outputs calculate spatial density near the wall.

\noindent
\textbf{Tracking.} The tracking module in the feeder, drinker, and activity pipelines associates chicken detections across frames to generate movement trajectories. Using a two-stage tracking-by-detection approach, the first stage groups detections into short tracklets based on proximity and appearance, improving efficiency with a greedy matching strategy over a sliding window. The second stage merges these tracklets into full trajectories using appearance features, motion continuity, and temporal gap constraints.

\noindent
\textbf{Feeding Behaviour Analysis.} This module, used in the feeder pipeline, quantifies feeding activity by analyzing how individual chickens interact with feeders over time. It uses outputs from the chicken detector, a feeder detection model based on YOLOX \cite{ge2021yolox} that identifies feeder locations, and the tracking module. Feeding events are identified when a chicken is positioned near a feeder, and are further refined using movement information to exclude short or transient interactions. These refined events are grouped into feeding bouts, allowing estimation of individual feeding duration and frequency across the video.

\noindent
\textbf{Drinking Behaviour Analysis.} This module, used in the drinker pipeline, quantifies drinking activity by analyzing the spatial relationship between chickens and drinker lines over time. It uses outputs from the chicken detector, a drinker line detection model based on YOLOX \cite{ge2021yolox}, and the tracking module. For each chicken, the distance from its center point to the nearest drinker line is calculated. If this distance falls below a defined threshold, the chicken is classified as drinking. This spatial proximity approach enables identification of individual drinking events throughout the video.

\noindent
\textbf{Clustering.} This module is used in the activity pipeline to analyze how chickens aggregate spatially within the activity area. It operates on the output of the tracking module, mapping chicken positions onto a spatial grid to generate density heatmaps over time. Clusters are identified by thresholding these heatmaps and applying connected component analysis to detect areas of local aggregation. The resulting metrics provide a way to quantify flock clustering behavior, which is used to study social dynamics and space use.

\noindent
\textbf{Density, weight, and activity analysis} are derived from detection and tracking outputs. Density is calculated by counting birds within a defined region of interest. Weight estimation uses the body width of feeding birds, adjusted for age, as input to a regression model. Activity levels are classified using motion features from trajectories, combined with age, to assign states like inactive, active, or highly active.

\subsection{Implementation}
The video analytics pipeline is implemented in Python 3.10 and uses various libraries to support the analytics task, such as PyTorch version 2.8.0, OpenCV version 4.7.0.72, scikit-image version 0.19.3, and ONNX Runtime GPU version 1.15.0. While libraries use GPU acceleration, the overall pipeline is a sequential process.

\section{Optimization Techniques}
\label{sec:optimisations}

This section outlines the system-level optimizations applied across the FlockFocus pipeline to improve runtime performance, reduce resource usage, and enable cost-efficient scaling across deployment settings. We present the optimizations and report their effect on performance alongside. Table~\ref{tab:opetimization_techniques} summarizes the optimization types and the modules they are applied to. All experiments were conducted on an AWS \texttt{g4dn.2xlarge} instance equipped with an NVIDIA T4 GPU (16\,GB VRAM), 8 vCPUs, 32\,GB of system memory, and a high-throughput NVMe SSD.

\begin{table}[t]
\centering
\small
\caption{Summary of Optimization Techniques: \#1: parallelism; \#2 libraries; \#3: video input.}
\begin{tabular}{|l|l|l|}
\hline
\textbf{Type} & \textbf{Optimization} & \textbf{Module} \\
\hline
\#1 & Parallel object detection (Section~\ref{sec:chunk-parallel}) & Detection \\
\#1 & Parallel object tracking (Section~\ref{sec:tracklet-parallelism}) & Tracking \\
\#1 & Object-level parallelism (Section~\ref{sec:feeding-behavior}) & Analysis \\
\#2 & Batched GPU inference (Section~\ref{sec:batch-inference}) & Detection \\
\#2 & Efficient post-processing (Section~\ref{sec:post-processing}) & Detection \\
\#2 & Code restructuring (Section~\ref{sec:clustering}) & Analysis \\
\#3 & Video formats (Section~\ref{sec:fmt}) & Video input \\
\#3 & Frame skipping (Section~\ref{sec:skip}) & Video input \\
\hline
\end{tabular}
\label{tab:opetimization_techniques}
\end{table}

\subsection{Batched GPU Inference for Detection}
\label{sec:batch-inference}
\noindent
\textbf{Performance challenge:}
GPU kernel invocations contain insufficient work to saturate the massively parallel computational resources provided by GPUs. Grouping together work from multiple kernel invocations increases GPU utilization.

\noindent
\textbf{Solution:}
We implement batched GPU inference to increase the amount of work per GPU kernel invocation. Instead of running inference frame by frame, the model processes a batch of frames in a single pass, reducing per-frame overhead and better utilizing GPU compute resources. This strategy is widely used in modern inference systems to improve efficiency and reduce cold start delays~\cite{ali2020batch, romero2021infaas, liu2023resource, zhang2019mark}. Batching increases efficiency of both data transfer and computation.

\begin{figure}[t]
  \centering
  \includegraphics[width=\linewidth]{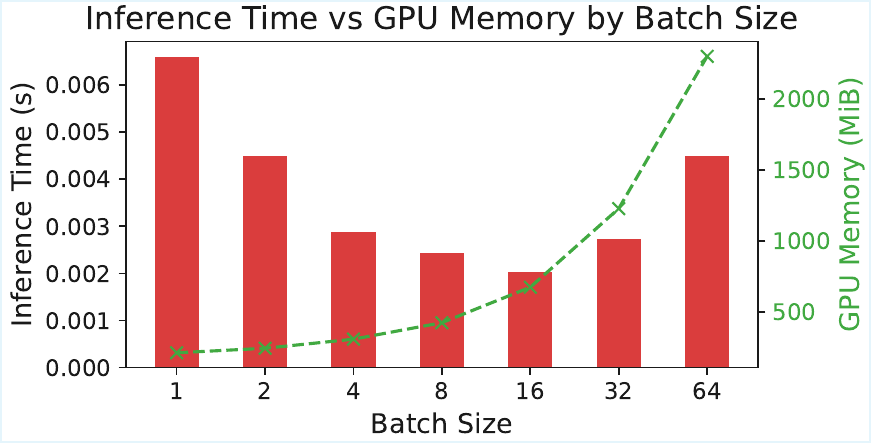}
  \caption{Effect of batch size on detection runtime and GPU memory utilization.}
  \Description{Detection time drops as batch size increases up to 16, while GPU memory use rises steadily.}
  
  \label{fig:batch-scaling}
\end{figure}

\noindent
\textbf{Evaluation:} We benchmarked the impact of different batch sizes on runtime and GPU memory usage, ranging from 1 to 64. As shown in Figure~\ref{fig:batch-scaling}, batch sizes up to 16 reduce inference time per frame from approximately 6.6~ms to 2~ms, while memory usage gradually increases and peaks at 2301~MiB. Beyond batch size 16, inference time increases due to memory pressure and diminishing returns.

\begin{figure}[t]
  \centering
  \includegraphics[width=0.8\linewidth]{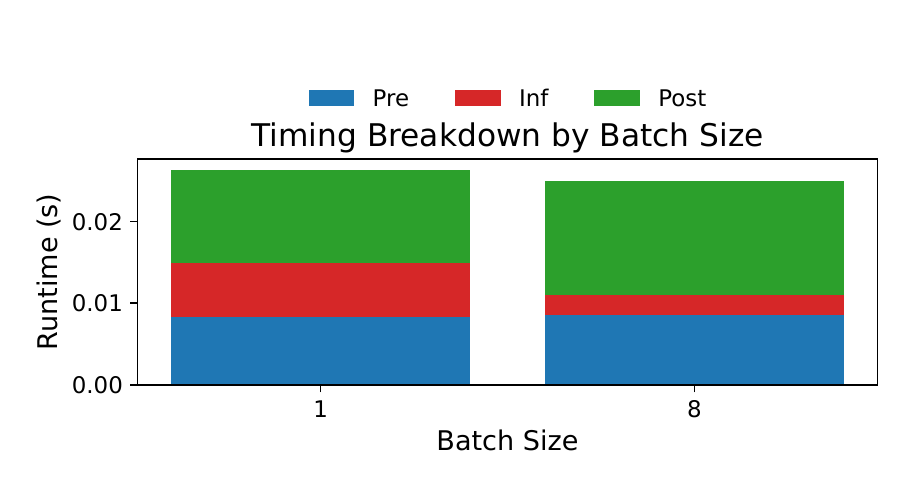}
  \caption{Breakdown of object detection module with U-Net before and after batching. While inference benefits from batching, the overall effect is obscured by costly preprocessing and post-processing steps.}

  \Description{After batching, inference time shrinks but preprocessing and post-processing dominate total latency.}
  \label{fig:unet-breakdown}
\end{figure}

\subsection{Accelerating Post-processing} \label{sec:post-processing}
\noindent
\textbf{Performance challenge:}
The performance benefits of batching are partially masked by the 
pre- and post-processing steps that surround the batched model inference step.
Batching lowers the inference time to under 5 ms per frame for a batch size of 8, but the combined time for pre-processing and post-processing still exceeds 15 ms per frame (see Figure~\ref{fig:unet-breakdown}).

The original implementation of the post-processing step relies on \texttt{skimage.regionprops} to group pixels into bounding boxes. This method is predominantly implemented in Python, and thus interpreted code. Neural network outputs were generated on GPU, and thus need to be transferred to CPU prior to post-processing. The data flow between GPU code and CPU code is affected by batching, which reduces locality and communication/computation overlap.

\noindent
\textbf{Solution:}
We replace \texttt{skimage.regionprops} with the OpenCV library’s \texttt{connectedComponentsWithStats}, 
a faster and more CPU-efficient method for extracting connected regions from the model outputs. 
This is coupled with a restructuring of the bounding box extraction logic to reduce memory overhead 
and reuse intermediate GPU tensors.

\begin{figure}[t]
  \centering
  \includegraphics[width=0.8\linewidth]{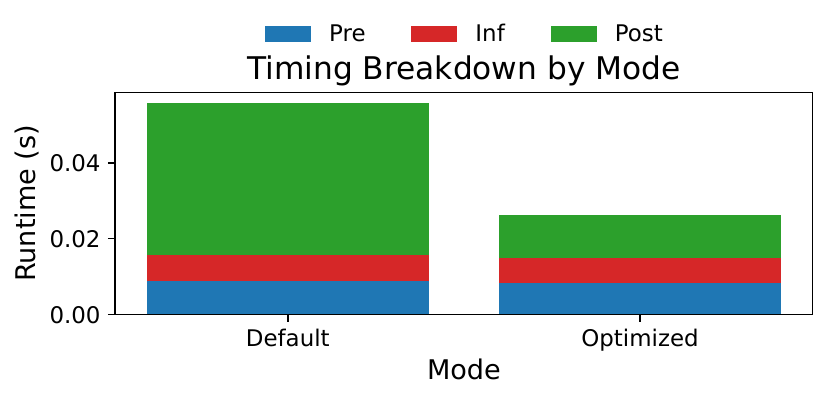}
  \caption{Breakdown of detection latency into preprocessing, inference, and post-processing components, before and after optimization.} 
  \Description{Optimized post-processing sharply reduces its share of total detection latency.}
  \label{fig:postprocessing-breakdown}
\end{figure}

\noindent
\textbf{Evaluation:} These modifications reduced post-processing time from 40.2 ms to 11.4 ms per frame. Figure~\ref{fig:postprocessing-breakdown} shows the breakdown of detection latency. In the default pipeline, the total cost was 55.8 ms per frame, consisting of 8.8 ms in preprocessing, 6.9 ms in inference, and 40.2 ms in post-processing. After optimization, the total dropped to 26.3 ms, with post-processing contributing only 11.4 ms. The changes made post-processing comparable in cost to the other components, and the bottleneck was effectively addressed.

 \subsection{Parallel Object Detection}
 \label{sec:chunk-parallel}
\noindent
\textbf{Performance challenge:} 
Object detection consists of many small and short-lived GPU kernels interspersed with idle gaps between tasks, leading to poor utilization of both GPU and CPU resources. For example, under the batch size of 8, GPU utilization remains below 20\%, whereas low GPU memory usage under 400~MiB shows scope for placing more work on the GPU.

\noindent
\textbf{Solution:}
There is inherent parallelism in the object detection task across frames, as objects are detected in each frame independently of other frames. We parallelize the object detection module by splitting the video stream in fixed-sized chunks of consecutive frames.

These chunks are assigned to separate worker processes via Python’s \texttt{multiprocessing} module. Each worker independently loads the detection model and runs inference on its assigned frames. This setup results in time-sharing of the GPU across processes. Idle time between kernels of one process are filled by kernel executions for other processes.

\begin{figure}[t]
  \centering
  \includegraphics[width=\linewidth]{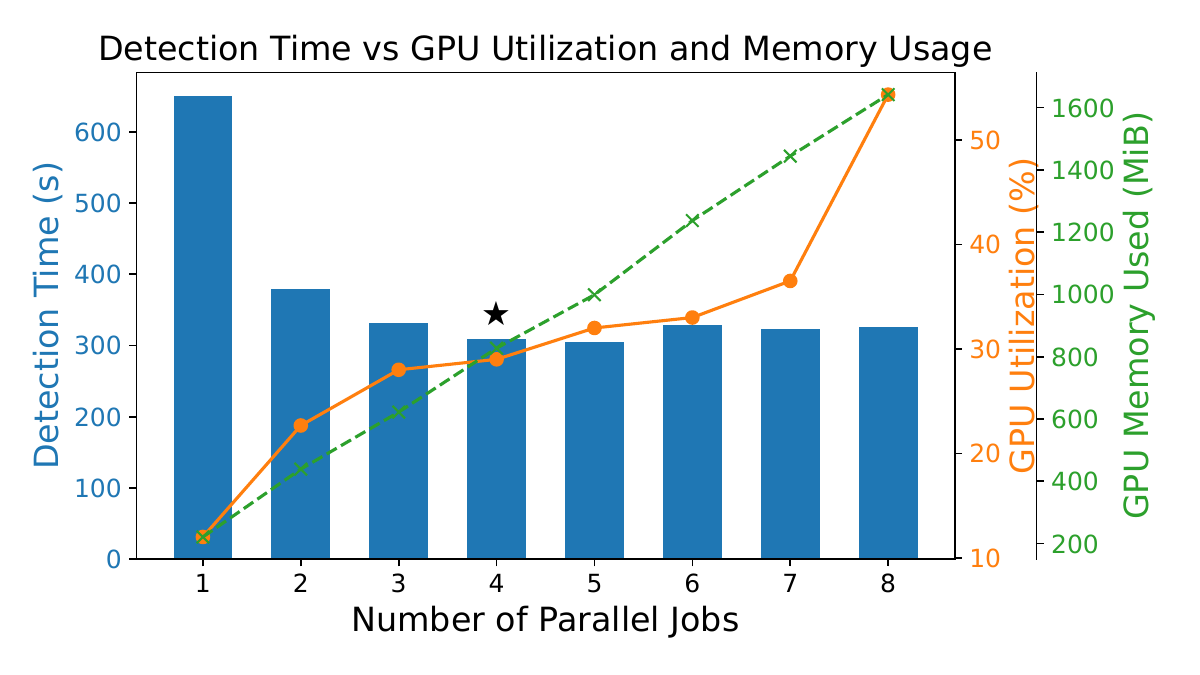}
  \caption{Detection time, GPU utilization, and GPU memory usage for varying numbers of parallel jobs. Speedup saturates after 4 workers due to increasing GPU resource contention.}

  \Description{Increasing parallel jobs lowers detection time until about four workers, then saturates due to GPU contention.}
  \label{fig:detection-parallel-scaling}
\end{figure}

\noindent
\textbf{Evaluation:}
This strategy increases GPU utilization and hides idle time by overlapping computation and I/O across processes, reduces scheduling delays and better utilizes memory bandwidth. In experiments with 1 to 8 parallel jobs (see Figure~\ref{fig:detection-parallel-scaling}), detection time decreases by 47\% when moving from 1 to 4 processes, while GPU utilization rose from around 20\% to over 50\%. However, additional gains beyond 4 jobs were marginal, with GPU memory usage saturating at around 1600~MiB and overall throughput plateauing.

\subsection{Parallel Object Tracking}
\label{sec:tracklet-parallelism}
\noindent
\textbf{Performance challenge:}
Object tracking merges occurrences of (seemingly) identical objects in successive frames into a tracklet (a trajectory segment). It is inherently a sequential process. Traditional tracking approaches, such as global network-flow solvers~\cite{zhang2008global}, are inefficient for long, high-resolution videos due to their sequential nature and high memory demands.

\noindent
\textbf{Solution:}
The original pipeline applies a two-stage tracker design~\cite{McLaughlin2015} for efficiency reasons. The first stage operates on fixed-size temporal windows (e.g., 5000 frames). It uses a greedy association strategy that combines bounding box overlap, motion continuity, and appearance features. The second stage is responsible for merging corresponding tracklets across windows. It links tracklets sequentially using cost-based association.

We introduce parallel execution in the first stage of the tracker, i.e., processing windows of frames independently and concurrently using \texttt{joblib.Parallel}.
However, the choice of window size affects tracking accuracy: shorter windows improve parallelism but can lead to fragmentation of tracklets and less accurate behavior analysis, while longer windows increase latency and memory usage.
As such, this optimization involves a speed vs.\ accuracy trade-off.

\noindent
\textbf{Evaluation:}
We measure tracking accuracy using two key metrics.
The first is \textit{average tracklet length}, defined as $L = \frac{N_\text{assigned}}{N_\text{tracklets}}$, where $N_\text{assigned}$ is the number of detections assigned to tracklets and $N_\text{tracks}$ is the number of unique object IDs. The second is \textit{detection coverage}, given by $C = \frac{N_\text{assigned}}{N_\text{total}}$, where $N_\text{total}$ is the total number of detections. Together, these metrics reflect the continuity and completeness of object identity assignment over time without requiring expensive manual labeling of all chickens in the video. 

\begin{figure}[t]
  \centering
  \includegraphics[width=\linewidth]{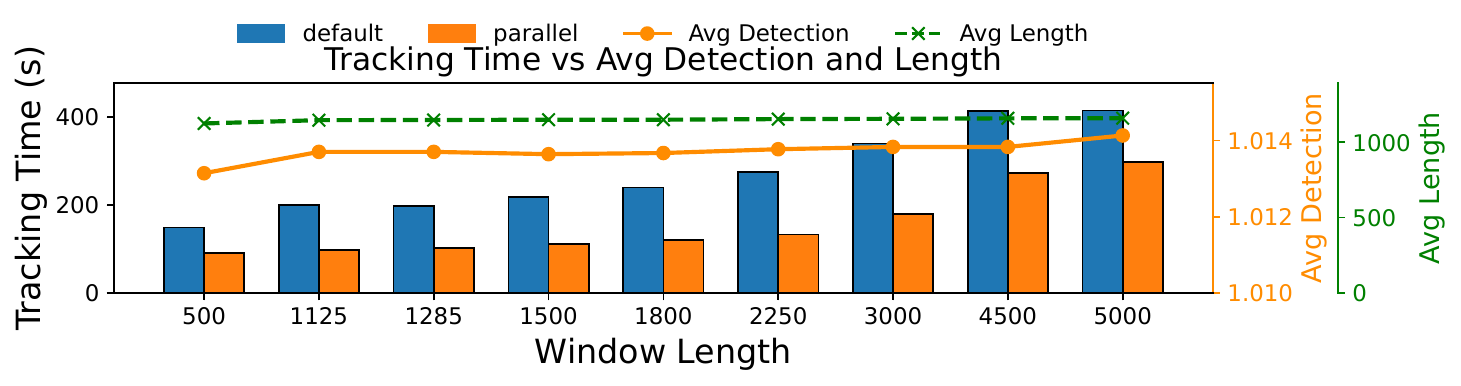}
  \caption{Speedup achieved by two-stage tracklet generation compared to the original sequential tracking pipeline.}
  
  \Description{Smaller tracking windows speed up processing with little change in tracklet length and minor coverage loss.}
  \label{fig:tracking-speedup}
\end{figure}

Figure~\ref{fig:tracking-speedup} characterizes the impact of window size on tracking time (bars) and the metrics (lines).
The window size is varied between 500 and 5000 frames over which average tracklet length ($L$) remains highly stable and detection coverage ($C$) decreases slowly with shorter window sizes.
The two-stage tracker design sees enhanced performance with reduced window size also in serial execution mode as its algorithmic complexity grows with window size.

The parallel tracker design achieves a 2.32× speedup over the original serial tracker. With a window size of 3000 frames, the parallel version runs in 178 s compared to 414 s for the serial version with a window size of 5000 frames, highlighting 3000 as the sweet spot in the speed–accuracy trade-off.

\subsection{Object-Level Parallelism}
\label{sec:feeding-behavior}
\noindent
\textbf{Performance challenge:}
Behavior analysis is implemented predominantly as a CPU algorithm as it involves machine learning algorithms that are not as compute-intensive as deep learning. As such, execution latency is high, and CPU and GPU utilization is low during behavior analysis.

\noindent
\textbf{Solution (locomotion and feeder pipelines):} 
We apply parallel execution on multiple CPU cores, making use of the observation that individual objects are processed independently. Each object is defined by the object tracker and corresponds to one chicken tracked across a number of video frames. Parallelism across objects is abundant, proportional to the number of identified chickens.
We partition chickens into $K$ groups, where $K$ is the number of parallel workers, and process each group in parallel using Python’s \texttt{multiprocessing.Pool}. The results are merged and sorted to respect the semantics of sequential execution order. 

\begin{figure}[t]
  \centering
  \includegraphics[width=\linewidth]{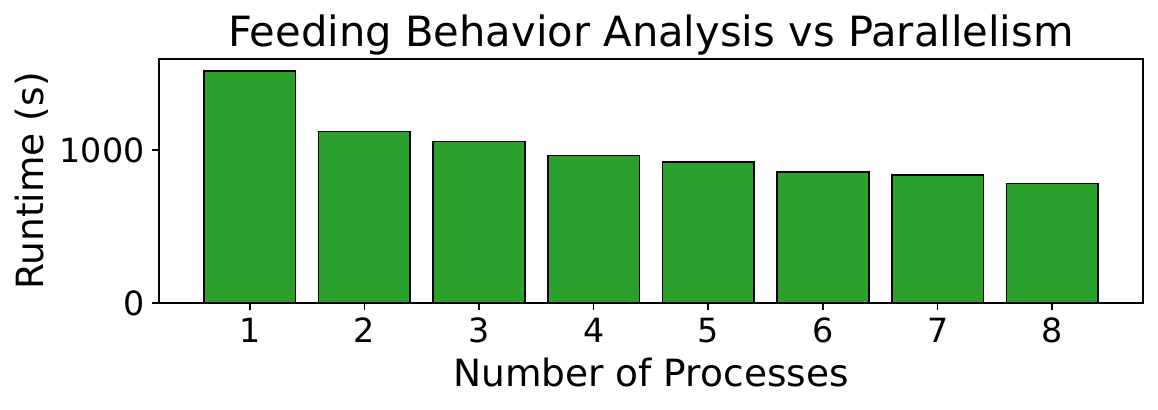}
  \caption{Runtime improvement from per-chicken parallelism in feeding behavior analysis.}

  \Description{Per-chicken parallelism significantly reduces runtime of feeding behaviour analysis.}
  \label{fig:chicken-parallelism}
\end{figure}

\noindent
\textbf{Evaluation:}
Parallelizing across detected chickens leverages unused CPU resources and reduces behavior analysis time from 1518$\:$s to 780$\:$s on typical farm videos with dozens of tracked chickens (Figure~\ref{fig:chicken-parallelism}).

\subsection{Efficient Clustering and Density Estimation}
\label{sec:clustering}

\begin{algorithm}[t]
    \footnotesize
    \caption{Default vs Optimized Clustering Pipeline}
    \label{alg:clustering}
    \begin{algorithmic}[1]
        \Procedure{DefaultClustering}{frames}
            \ForAll{frames}
                \State Filter chickens to ROI
                \State Initialize grid
                \ForAll{grid cells}
                    \ForAll{chickens}
                        \State Compute distance and update density
                    \EndFor
                \EndFor
                \State Compute heatmap, entropy, clustering degree
            \EndFor
        \EndProcedure
        \Statex

        \setcounter{ALG@line}{0}
        \Procedure{OptimizedClustering}{frames}
            \State Filter ROI once; initialize grid once \Comment{ROI/grid reuse}
            \State Partition frames into batches
            \ForAll{batches \textbf{in parallel}} \Comment{Parallelization}
                \State Vectorized density map (\texttt{cdist}) \Comment{Vectorization}
                \State Compute heatmap, entropy, clustering degree
            \EndFor
        \EndProcedure
    \end{algorithmic}
\end{algorithm}

\noindent \textbf{Performance Challenge:}  
Spatial clustering of chickens provides key insight into flock behavior, including rest patterns, social grouping, and crowding-induced stress. This analysis requires computing frame-level density maps using distance, entropy, and clustering degree from tracked chicken positions. However, the original implementation relied on nested loops for distance computation and region analysis, which became a severe performance bottleneck in long videos.  

\noindent \textbf{Solution:}  
To address this bottleneck, we introduced three key optimizations, summarized in Algorithm~\ref{alg:clustering}:  
1) \emph{Vectorized distance computation} using \texttt{scipy.spatial.distance.cdist} to replace loop-based grid-to-bird distance calculations that compose the density map.  
2) \emph{Batch-wise parallelization} of frame-level clustering statistics using \texttt{joblib.Parallel} to exploit multi-core CPUs.  
3) \emph{Heatmap and ROI optimizations} through streamlined density map normalization, grid initialization, and one-time ROI filtering to reduce redundancy.  

\begin{figure}[t]
  \centering
  \includegraphics[width=0.7\linewidth]{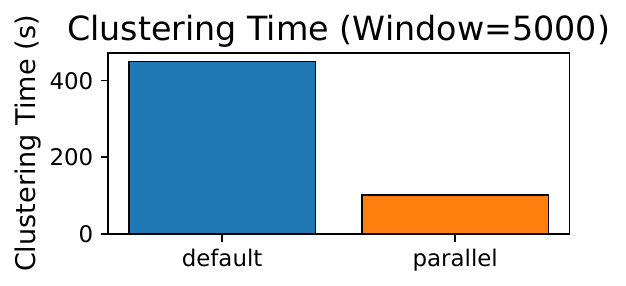}
  \caption{Improvements for clustering and density estimation after vectorization and parallelization.}

  \Description{Vectorization and parallelization make clustering and density estimation much faster.}
  \label{fig:clustering-speedup}
\end{figure}

\noindent \textbf{Evaluation:}  
These optimizations reduced clustering runtime from approximately 449 seconds to 101 seconds per video (Figure~\ref{fig:clustering-speedup}), while preserving analytical consistency with the original implementation. This efficiency enables integration of clustering analytics into near real-time monitoring pipelines and scalable cloud deployments.

\subsection{Impact of Video Formats}
\label{sec:fmt}
\noindent \textbf{Performance Challenge:}  
Video ingestion can significantly impact the efficiency of the clustering pipeline. Frame loading performance, accuracy, and downstream efficiency depend on both video format and resolution. The key question is whether higher decoding throughput can be achieved without compromising detection accuracy. 

\noindent 
\textbf{Solution:}  
To address this, we conducted an empirical study using the formats natively supported by our camera: H.264, H.265/HEVC, and MJPEG. Each format was evaluated across multiple resolutions to capture the trade-offs between speed and accuracy. We measured both raw frame loading throughput (FPS) and downstream accuracy metrics, including average tracklet length and detection coverage.  

\begin{figure}[t]
  \centering
  \includegraphics[width=\linewidth]{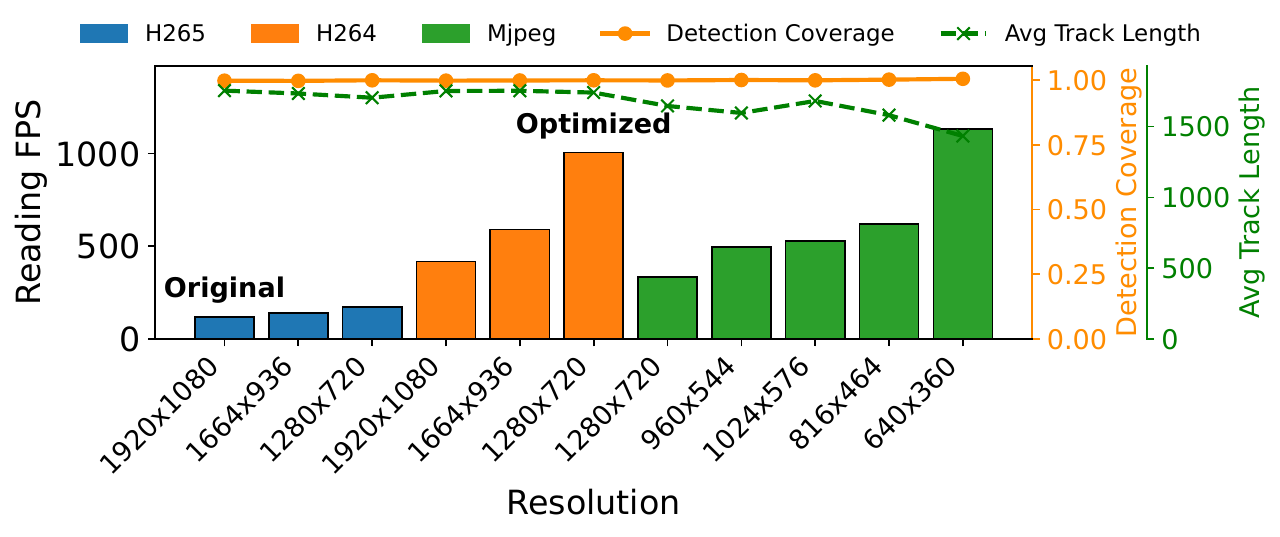}
  \caption{Reading throughput by codec and resolution. The optimized path (selective decoding plus lower resolutions) delivers the highest ingestion rates, with MJPEG scaling best and exceeding 3{,}000 FPS at 480p; H.264/H.265 remain lower at the same resolutions.}

  \Description{Reading throughput improves at lower resolutions; MJPEG is fastest, especially at 480p.}
  
  \label{fig:read-fps}
\end{figure}

\noindent \textbf{Evaluation:}  
Figure~\ref{fig:read-fps} shows that reading FPS rises sharply as resolution decreases and codec choice plays a decisive role. MJPEG achieved the highest throughput, exceeding 3{,}000 FPS at 480p, while H.264 and H.265 remained consistently lower at the same resolution. However, MJPEG’s intra-frame encoding introduced subtle artifacts in low-light or occluded regions, slightly affecting segmentation and detection robustness. By contrast, reducing resolution from 1080p to 720p in H.264 preserved accuracy while reducing decoding latency by more than $2\times$. Based on this trade-off, we selected H.264 at 720p as the default configuration, as it balances speed and robustness/accuracy without compromising analysis.

\begin{figure}[t]
  \centering
  \includegraphics[width=\linewidth]{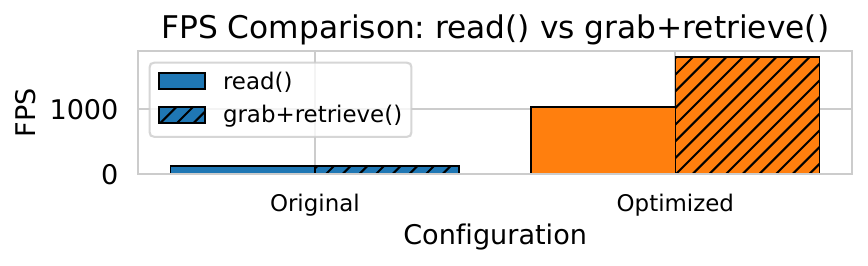}
  \caption{Frame access time comparison between \texttt{read() and \texttt{grab()+retrieve()} for different fps on selected set of videos}}

  \Description{Using grab+retrieve cuts frame access time compared to read across frame rates.}
  
  \label{fig:grab-retrieve}
\end{figure}

\subsection{Frame Skipping}
\label{sec:skip}
\noindent \textbf{Performance Challenge:} 
In the conventional \texttt{cv2.read()} workflow, every frame is fully decoded regardless of whether it is used.
The analytics pipeline samples 2 FPS from a 30 FPS video. This way, 14 out of every 15 frames are decoded and then discarded, wasting CPU cycles and memory bandwidth.

\noindent \textbf{Solution:}
We implemented OpenCV’s two-step interface. The \texttt{grab()} call advances the stream pointer without decoding, while \texttt{retrieve()} is invoked only for frames that are processed. This avoids redundant decoding work while maintaining temporal consistency in the frame stream.

\noindent \textbf{Evaluation:}
The grab+retrieve strategy yields substantial improvements in throughput and lowers CPU load (Figure~\ref{fig:grab-retrieve}). When combined with the codec and resolution selection described above, this optimization provides a complementary reduction in video ingestion overhead, enabling faster and more scalable downstream analysis. We observe that the video encoding format used impacts on the strength of this optimization.

\section{End to End Evaluation}
\label{sec:evaluation}

\begin{figure}[t]
  \centering
  \includegraphics[width=0.9\linewidth]{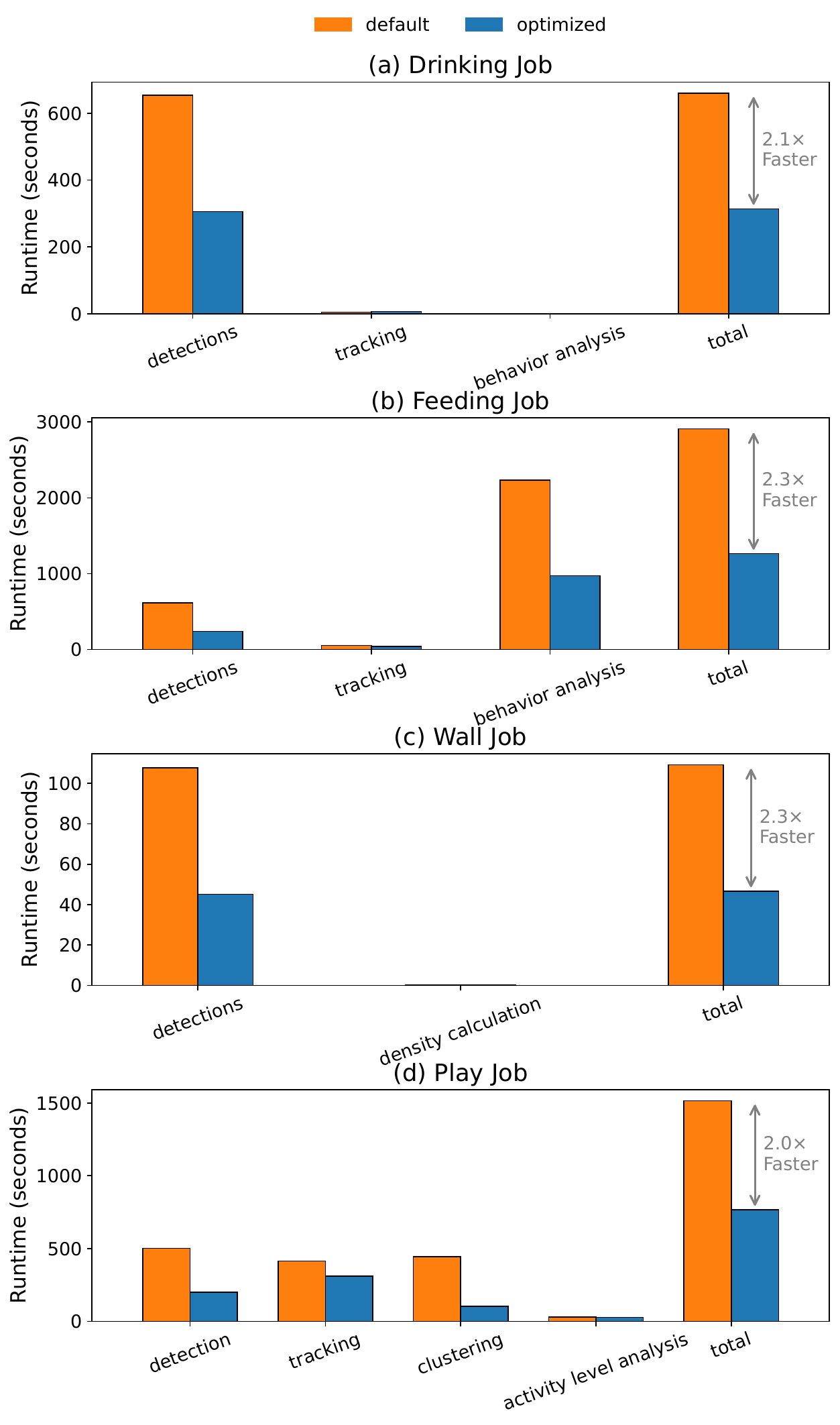}
  \caption{End-to-end runtime comparison of default vs. optimized pipelines for each zone. Optimizations across detection, tracking, and analysis modules yielded 2.0–2.3× speedups.}

  \Description{Across all four pipelines, the optimized version halves end-to-end runtime compared to default.}
  
  \label{fig:e2e}
\end{figure}

\begin{figure}[t]
  \centering
  \includegraphics[width=\linewidth]{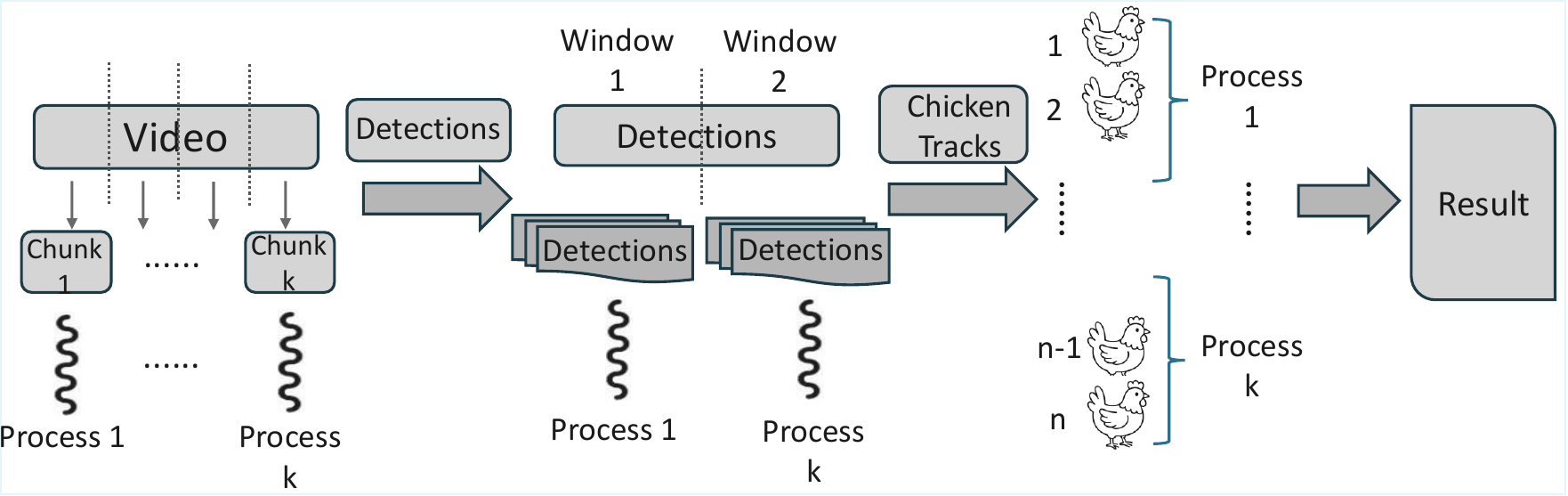}
  \caption{Parallelism in the analytics pipelines.}
  
  \Description{Diagram showing chunk and window parallelism with results merged into detections and tracks.}
  \label{fig:parallelism}
\end{figure}

We evaluated the end-to-end performance of the optimized inference pipelines on real-world farm videos (5 minutes at 30 FPS, $\sim$9000 frames) across four behavior monitoring zones. Figure~\ref{fig:parallelism} summarizes the parallelism introduced in the analytics pipelines. The semantics of the analytics require processing the whole video at once. Modules are parallelized along dimensions of the data, implying a full instantiation of the intermediate data between any two pipeline modules.

Figure~\ref{fig:e2e}a-d shows runtime improvements for each pipeline. The \textbf{drinker pipeline} (Figure~\ref{fig:e2e}a) achieved a 2.1$\times$ speedup, reducing runtime from 660 to 314 seconds through chunk-based detection parallelism (Section~\ref{sec:chunk-parallel}), batched GPU inference (Section~\ref{sec:batch-inference}), efficient post-processing (Section~\ref{sec:post-processing}), and tracklet-level parallelism in tracking (Section~\ref{sec:tracklet-parallelism}). The \textbf{feeding pipeline} (Figure~\ref{fig:e2e}b), the most complex, used the same detection optimizations, windowed parallelism in tracking, and per-chicken multiprocessing with clustering vectorization in behavior inference (Section~\ref{sec:feeding-behavior}), cutting runtime from 29908 to 1263 seconds (2.3$\times$). The \textbf{wall pipeline} (Figure~\ref{fig:e2e}c), which includes only detection and density estimation, dropped from 109 to 46 seconds (2.3$\times$) with chunk-based detection and optimized post-processing. The \textbf{activity pipeline} (Figure~\ref{fig:e2e}d) showed a 2.0$\times$ speedup, from 1515 to 766 seconds, via detection improvements and vectorized clustering in behavior analysis (Section~\ref{sec:clustering}). Table~\ref{tab:e2e-cost} summarizes the financial impact, with processing cost per video reduced from \$0.691 to \$0.312 (54.8\% reduction).

\begin{table}[t]
\centering
\footnotesize
\caption{Cost Comparison of Default vs Optimized for a 5 min video processing}
\label{tab:e2e-cost}
\begin{tabular}{lcc}
\toprule
\textbf{Pipeline} & \textbf{Default Cost (\$)} & \textbf{Optimized Cost (\$)} \\
\midrule
Drinking          & 0.077  & 0.037 \\
Wall              & 0.014  & 0.006 \\
Feeding           & 0.404  & 0.171 \\
Activity          & 0.196  & 0.098 \\
\midrule
\textbf{Total}    & \textbf{0.691} & \textbf{0.312} \\
\bottomrule
\end{tabular}
\end{table}
\section{Discussion}

\begin{figure}[t]
\centering
\includegraphics[width=0.8\linewidth]{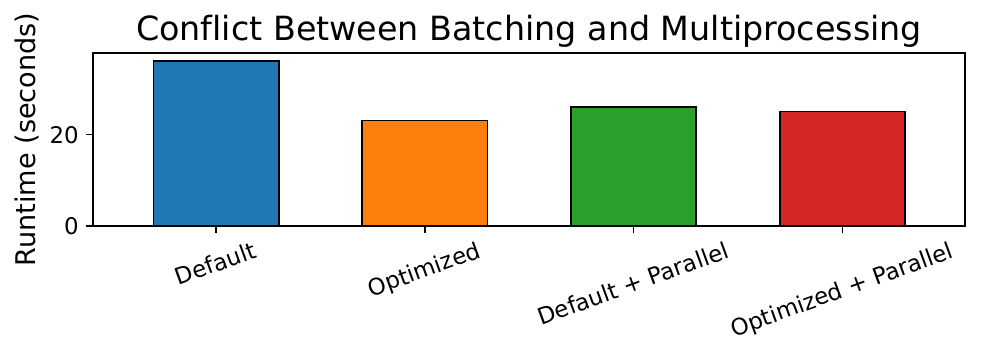}  
\caption{Detection runtime across configurations for the drinker job (500 frames). While batching and multiprocessing reduce latency, their 
yields limited additional speedup due to GPU contention.}

\Description{Batching and multiprocessing each help, but together give limited extra speedup due to GPU contention.}

\label{fig:batching-parallelism-conflict}
\end{figure}

Optimizing GPU modules in \textit{FlockFocus} revealed performance trade-offs not obvious when evaluating batching and multiprocessing separately. Batched inference reduced detection latency, but combining it with chunk-based multiprocessing caused resource contention, limiting further gains. Figure~\ref{fig:batching-parallelism-conflict} shows the legacy configuration (non-batched, single-process) took 36 seconds, while batched inference lowered it to 23 seconds. With multiprocessing, the time improved to 25 seconds. This suggests that batching and multiprocessing don’t combine cleanly, as larger batches increase GPU workloads, leading to memory contention. Coordinated strategies like CUDA streams~\cite{cuda-streams} or Multi-Process Service (MPS)~\cite{cuda-mps} have the potential to avoid blocking and competition for memory and compute bandwidth in long-duration workloads.

A similar bottleneck occurred in post-processing. Despite accelerated detection, steps like connected component analysis and bounding box extraction remained CPU-bound. Increasing parallel jobs raised CPU load, limiting speedup, especially in the wall and feeder pipelines, where speedup plateaued. Improving throughput requires more than adding workers or batching; the pipeline needs holistic optimization with careful scheduling, memory sharing, and data transfer. Further optimizations could include stream-level GPU scheduling, adaptive batch sizing, and overlapping pipeline modules to improve concurrency without bottlenecks.
\section{Related Work}
\label{sec:related-works}

\textbf{Cloud Inference Optimization}  
Recent work has explored how to reduce the cost and latency of video inference in cloud environments, particularly under high-throughput, multi-model workloads. LLama~\cite{romero2021llama} models video pipelines as directed acyclic graphs (DAGs) and explores configuration tuning across sampling, batching, and hardware use to optimize runtime efficiency. However, due to the combinatorial configuration space and varying data paths, manual tuning becomes infeasible at scale. Scrooge~\cite{hu2021scrooge} addresses this by efficiently packing workloads into GPU-equipped Virtual Machines (VMs), using optimization-based placement to meet performance targets while lowering serving costs. IPA~\cite{ghafouri2025ipa} introduces dynamic pipeline adaptation using model variants, integer programming, and SLA-driven configuration to balance latency, cost, and accuracy. However, model variant changes affect the accuracy of the inference pipeline. StreamBox~\cite{wu2024streambox} tackles startup overhead and inter-process communication costs in serverless GPU inference by enabling fine-grained memory sharing and efficient intra-GPU communication. Our work addresses similar challenges in batching, scheduling, and inter-stage I/O, but focuses on optimizing offline video analytics through a unified, efficient pipeline architecture.

\textbf{Edge Inference Optimization}  
Due to limited compute, memory, and bandwidth, edge environments require aggressive optimization of inference pipelines, and many of these techniques are equally relevant in the cloud. Turbo~\cite{lu2022turbo} opportunistically boosts inference quality using idle GPU cycles, while Kum et al.~\cite{kum2022optimization} optimize adaptive batching and model reuse based on workload shifts. OCTOPINF~\cite{nguyen2025octopinf} adapts job batching and scheduling in response to GPU workloads, while NVIDIA's DeepStream~\cite{nvidia2023deepstream} provides a practical baseline for modular edge inference. Gemel~\cite{padmanabhan2023gemel} reduces GPU memory usage by merging layers across related models, improving inference efficiency without sacrificing accuracy. Cakic et al.~\cite{cakic2023developing} deployed detection and segmentation models on edge devices for poultry monitoring, and Srinivasagan et al.~\cite{srinivasagan2025edge} showed that TinyML can classify chicken vocalizations with high accuracy on low-power devices.

\section{Conclusion}

\textit{FlockFocus} was developed for a near real-time, behavior-aware video analysis across poultry zones using shared cloud resources. This work presented system-level optimizations that significantly improved throughput and reduced latency, achieving up to $2.3\times$ performance gains through batched inference, chunk-based multiprocessing, and parallel post-processing. These optimizations enabled more efficient use of computational resources, reducing processing times for key modules and ensuring high-performance video analytics in poultry welfare monitoring. Future improvements will focus on further enhancing GPU concurrency, exploring streaming detections into tracking, dynamic batching, selective frame skipping, and overlapping pipeline stages to reduce idle time and maximize resource utilization for real-time deployment in large-scale farm settings.

\begin{acks}

The research reported in this publication was supported by the Foundation for Food \& Agriculture Research (FFAR) and McDonald’s Corporation under the SMART Broiler Programme (Grant ID: 21-000053). 

\end{acks}

\bibliographystyle{ACM-Reference-Format}
\bibliography{references}

\end{document}